\documentclass[letterpaper]{article}

\usepackage{parskip}
\usepackage{graphicx}
\usepackage{amsmath}
\usepackage{caption}
\usepackage{chngpage}
\usepackage{amssymb}
\usepackage{textgreek}

\begin{document}
\title{How Unique is Milwaukee’s 53206?\\ 
An Examination of Disaggregated Socioeconomic Characteristics Across the City and Beyond
}
\author{Scott W. Hegerty\\
Department of Economics\\
Northeastern Illinois University\\
Chicago, IL 60625\\S-Hegerty@neiu.edu
}
\date{\ March 7, 2020}
\maketitle

\begin{abstract}

Milwaukee’s 53206 ZIP code, located on the city’s near North Side, has drawn considerable attention for its poverty and incarceration rates, as well as for its large proportion of vacant properties. As a result, it has benefited from targeted policies at the city level. Keeping in mind that ZIP codes are often not the most effective unit of geographic analysis, this study investigates Milwaukee’s socioeconomic conditions at the block group level. These smaller areas’ statistics are then compared with those of their corresponding ZIP codes. The 53206 ZIP code is compared against others in Milwaukee for eight socioeconomic variables and is found to be near the extreme end of most rankings. This ZIP code would also be among Chicago’s most extreme areas, but would lie near the middle of the rankings if located in Detroit. Parts of other ZIP codes, which are often adjacent, are statistically similar to 53206, however—suggesting that a focus solely on ZIP codes, while a convenient shorthand, might overlook neighborhoods that have similar need for investment. A multivariate index created for this study performs similarly to a standard multivariate index of economic deprivation if spatial correlation is taken into account, confirming that poverty and other socioeconomic stresses are clustered, both in the 53206 ZIP code and across Milwaukee.   

\end{abstract}

\section{Introduction}
One of the lowest-income areas of the city of Milwaukee, the 53206 ZIP code has received a great deal of attention both locally and nationally. While the claim that this area has the highest incarceration rate in the country has been disputed, it nonetheless faces concentrated poverty and faces other economic stresses. A detailed analysis by Levine (2019) compares this area with the rest of Milwaukee and with other parts of the metro area, for a number of socioeconomic variables including employment, housing tenure, rent burdens, and insurance coverage. A documentary film by McQuirter (2016) delves into conditions in this ZIP code, particularly for incarcerated residents and their families. City leaders have directed resources into this area, notably to help improve childbirth outcomes\footnote{See Spicuzza, \textit{Milwaukee Journal Sentinel}, 4/2/2019, for example for Milwaukee’s efforts to bring doula services to the 53206.}. But while the 53206 ZIP code is clearly poorer than much of the rest of the Milwaukee metropolitan area, less is known about the concentration or dispersion of similarly-stressed areas of the city. 

In addition, a focus on the ZIP code as the unit of analysis faces the well-known “modifiable areal unit problem” (Openshaw, 1984), whereby results differ by the choice of the study area in question. Given that ZIP codes were introduced in the 1960s specifically for postal purposes, they would not be expected to be optimized for socioeconomic analysis. In fact, a Congressional Research Service report (Ginsberg, 2011), notes that ZIP code boundaries are often problematic when used for purposes such as calculating insurance rates. There is need, therefore, to go beyond this unit of analysis when examining urban social issues. 

This study examines the 551 block groups in the City of Milwaukee, in addition to the 27 ZIP codes that are entirely or partially within the city. This will allow for a finer analysis that can isolate economically distressed neighborhoods alongside better-performing ones. It also allows us to focus entirely on the city itself, rather than on suburban tracts. This study has three goals. First, it examines distributions of block groups within each ZIP code, to highlight the variation within these larger areas and to compare parts of each Milwaukee ZIP code to the 53206 median value. Second, we create a multivariate index to capture socioeconomic conditions across the Milwaukee, as well as for the comparison cities of Detroit and Chicago. Finally, we compare this index with a separate measure of economic deprivation to test whether our new index provides any useful information regarding the socioeconomic makeup of the 53206 ZIP code, as well as the rest of Milwaukee.  

\begin{figure}[ht]
\hfill
\caption{Poverty Rates at the ZIP-Code and Census Block-Group Levels, 2016.}
\includegraphics[width=1\textwidth]{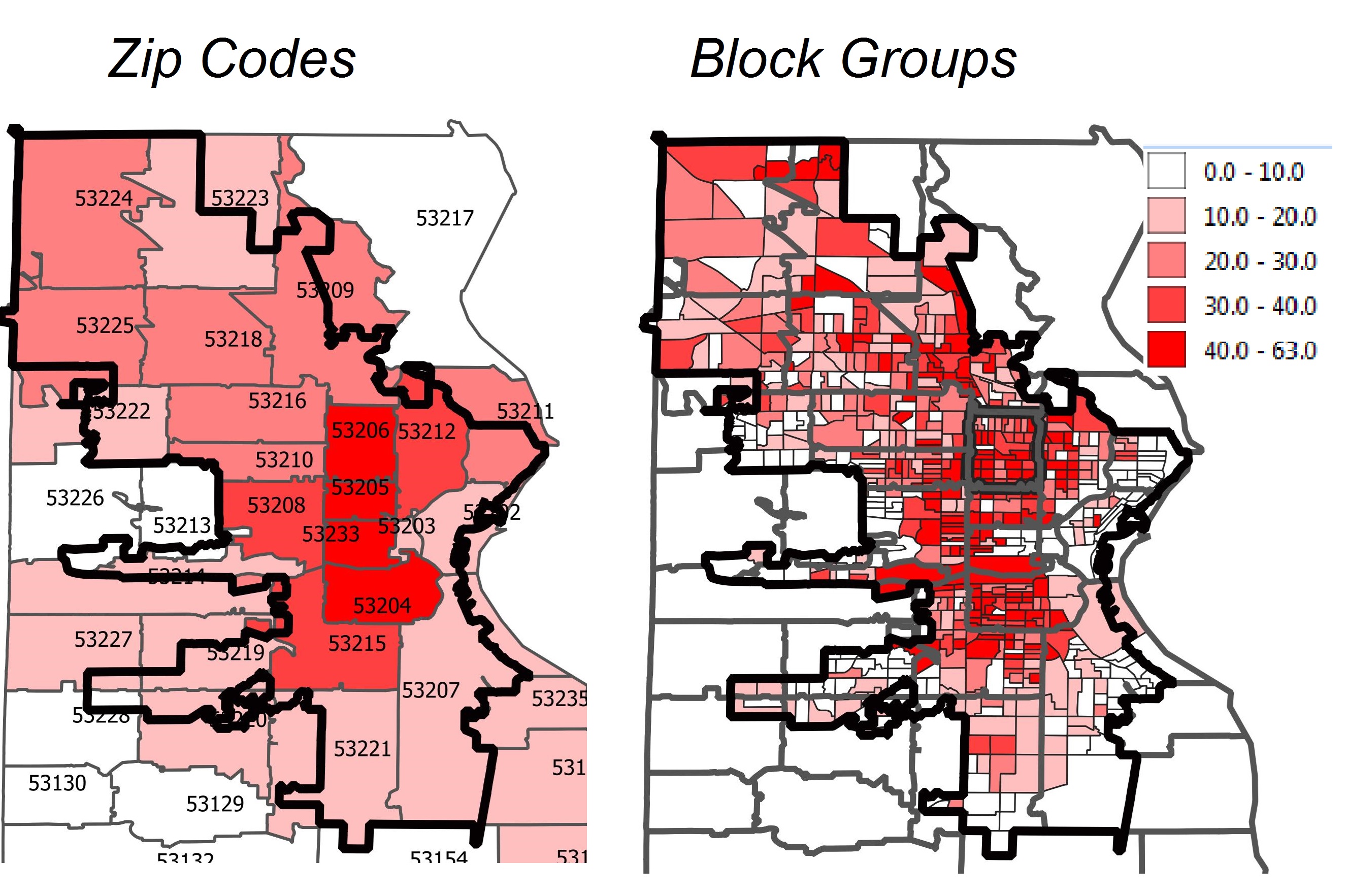}
\end{figure}

Figure 1 depicts Milwaukee’s poverty rates by ZIP code, as well as by block groups. The highest-poverty ZIP codes extend south from 53206 (encompassing 53205, 53233, and 53204). But as the block-group-level poverty rates show, there are areas of equally high poverty spread out across the city. This is particularly true for the near South Side and on the Northwest Side. We see high values in the 53208, 53210, and 53212 ZIP codes, as well as the Milwaukee portion of 53209, which is shared with the suburb of Glendale.

It is these variations—pockets of high poverty or urban distress alongside better-performing areas, leading to an inaccurate aggregate measure—that is one of the main areas of focus in this study. We also question whether poverty alone is the most informative measure of conditions across the city of Milwaukee. To address these issues, we examine a set of variables including economic distress, ethnicity, and the housing stock, for the Census block groups within each ZIP code\footnote{For further discussion on poverty and economic segregation, particularly in urban areas in the United States, see Jargowsky (1996), Wilson (2008-2009), and for a discussion of relevant quantitative methods, see Massey and Denton (1988).}.  We find that the average (median) values for the constituent block groups are often—but not always—close to the aggregate ZIP code value. In addition, the top 25\% of block groups in other Milwaukee ZIP codes resemble the median 53206 values. This suggests that pockets of distress might be “missed” if policymakers focus only on larger statistical areas.     

We also create a multivariate “53206” index, which captures five of our variables, to create a single measure that can be compared both across Milwaukee and with other cities. We find that, as expected, index values are highest on Milwaukee’s near North Side. When we re-create the index for block groups in Detroit and Chicago, we find that the 53206 would be close to the middle among a ranked list of Detroit ZIP codes, but near the top for Chicago.

Finally, we compare this “53206” index against a commonly used measure of economic deprivation. We also incorporate geography into our analysis by measuring spatial autocorrelation and applying spatially-weighted regression analysis. We find that the “53206” index captures neighborhood characteristics that the Deprivation index does not, but that the differences are reduced once spatial methods are applied. As such, we consider these findings to support that these measures are in fact spatially concentrated, both within this specific ZIP code, but also in neighborhoods elsewhere in the city. 

\begin{table}[ht]
\caption{Block Group and ZIP Code Population Statistics in Milwaukee.}
  \begin{adjustwidth}{-.5in}{-.5in}  
        \begin{center}
\begin{tabular}{lrrrrlrrrr}

ZIP&\# BGs&ZIP Pop.&BG Pop.&\% of Total&ZIP&\# BGs&ZIP Pop.&BG Pop.&\% of Total\\
\hline
53202&19&24830&25857&104.1&53215&45&61368&55552&90.5\\
53203&1&1263&1284&101.7&\textbf{53216}&35&33174&36228&109.2\\
53204&38&39895&39603&99.3&53218&43&41387&42432&102.5\\
\textbf{53205}&14&10309&13699&132.9&53219&21&34580&18066&52.2\\
\textbf{53206}&32&24805&22666&91.4&53220&10&26388&11036&41.8\\
53207&36&36917&35595&96.4&53221&23&40007&32109&80.3\\
53208&34&30979&32780&105.8&53222&23&26082&23544&90.3\\
\textbf{53209}&40&45814&34007&74.2&53223&14&28514&18883&66.2\\
\textbf{53210}&28&26683&22725&85.2&53224&13&22416&24260&108.2\\
53211&16&35414&19110&54.0&53225&18&26235&22074&84.1\\
\textbf{53212}&36&31788&31522&99.2&53226&2&19247&3117&16.2\\
53213&3&26360&2407&9.1&53227&2&23548&3780&16.1\\
53214&6&34752&6119&17.6&53228&2&14838&3552&23.9\\
 & & & & &53233&13&15786&15871&100.5\\

\hline
\end{tabular}
\end{center}
    \end{adjustwidth}
\caption*{Bold = adjacent to 53206.\\
 \% of Total = BG population divided by ZIP population.}
\end{table}

Table 1 compares the ZIP code areas with the totals of the block groups that are considered to comprise them. Those that lie completely within the city have values that are generally close to one another. One exception is the 53205 ZIP code, where the block groups within it have a larger combined population than the ZIP code area itself. The rest of the areas have values that are within about 10 percent of one another. The block groups within the 53206 ZIP code contain about 92 percent of the ZIP code’s population. In terms of population, 53206 ranks in the bottom half among city ZIP codes, with 53215 in first place. Because the constituent block groups are similar to their corresponding ZIP codes, we are able to use their distributions to compare each as a whole, as well as portions of each that might have extremely high or extremely low values. This will allow us to assess whether the 53206 ZIP code is in fact a special case, both locally as well as nationally.

\section{Methodology}
In this study, we use U.S. Census data (2016 American Community Survey 5-year estimates), for Milwaukee ZIP codes and block groups. The measures we derive (usually as percentages of the total) include the poverty rate among households, the percentage of households receiving SNAP benefits, the percentages of Black and White residents, the vacancy rate, the percentage of renters, the percentage of mortgage holders, and the percentage of buildings built before 1939. These capture economic conditions, ethnicity, and living or neighborhood conditions. In particular, housing tenure has been shown to influence crime or other neighborhood conditions (Lockwood, 2007, or Hegerty, 2017), as residents are less invested in the quality of their neighborhoods. On the other hand, homeownership rates vary by neighborhood, by city, and by country, so this relationship is not particularly straightforward. Likewise, low rates of mortgage holders might indicate a high degree of financial exclusion, but they might also be related to high proportions of renters or to low property values (which would allow for cash transactions). Finally, older parts of the city might have antiquated housing stock; this could imply a degree of deterioration, but need not necessarily be the case. We therefore use caution when interpreting any of our variables either positively or negatively.

\begin{table}[ht]
\caption{Spearman Correlation Coefficients.}
  \begin{adjustwidth}{-1.5in}{-1.5in}  
        \begin{center}
\begin{tabular}{lrrrrrrr}
&\%VAC&\%SNAP&\%RENT&\%WHITE&\%BLACK&\%POV&\%MORTG\\
\hline
PERC1939&0.334&0.324&0.297&-0.249&0.016&0.297&-0.082\\
PERCVAC&1&0.485&0.346&-0.471&0.400&0.391&-0.145\\
PERCSNAP&&1&0.546&-0.777&0.551&0.772&-0.217\\
PERCRENT&&&1&-0.351&0.302&0.592&-0.229\\
PERCWHITE&&&&1&-0.793&-0.644&0.155\\
PERCBLACK&&&&&1&0.463&-0.084\\
PERCPOV&&&&&&1&-0.246\\

\hline

\end{tabular}
        \end{center}
    \end{adjustwidth}
\end{table}

Table 2 presents the Spearman correlations between each pair of variables. The largest negative associations are among the percentages of White residents, SNAP recipients and Black residents. There is a large, positive association between proportions of SNAP recipients and poverty rates, but the correlation is not perfect. Old housing stock is not correlated with many of the other variables; is the percentage of households with mortgages has even lower values.

We also consider, but exclude, other potential variables. The unemployment rate is used (alongside an education variable) later in the study, but since it is often low if labor force participation is low, it sometimes fails to provide useful information. We also decide not to use any non-Census data (such as incarceration rates, crime rates, or proportions of vacant parcels), so that our measure can be compared across cities. As a result, we examine eight socioeconomic variables in detail, and use five of them in our “53206” index, detailed below. We create a second, multivariate “Deprivation” index using one of these variables and two additional variables that are commonly used in the literature. 

For those block groups that have their centroids within a particular ZIP code, we then calculate three quantile values. Our most important is the median, but we also consider the 25\% and 75\% values to capture pockets of low or high values within each area. This is particularly important for those ZIP codes with a low overall poverty rate, for example; poor block groups might not be captured by the aggregate, or even average, block group values. We note cases in which the ZIP code statistics differ from the median block group value. We also rank each ZIP code by median value, assessing where 53206 falls with respect to other parts of the city. The 25\% and 75\% values help show parts of other ZIP codes that resemble 53206 statistically.

\begin{figure}[p]
\hfill
\caption{Block-Group Distributions by ZIP Code.}

\includegraphics[width=.5\textwidth]{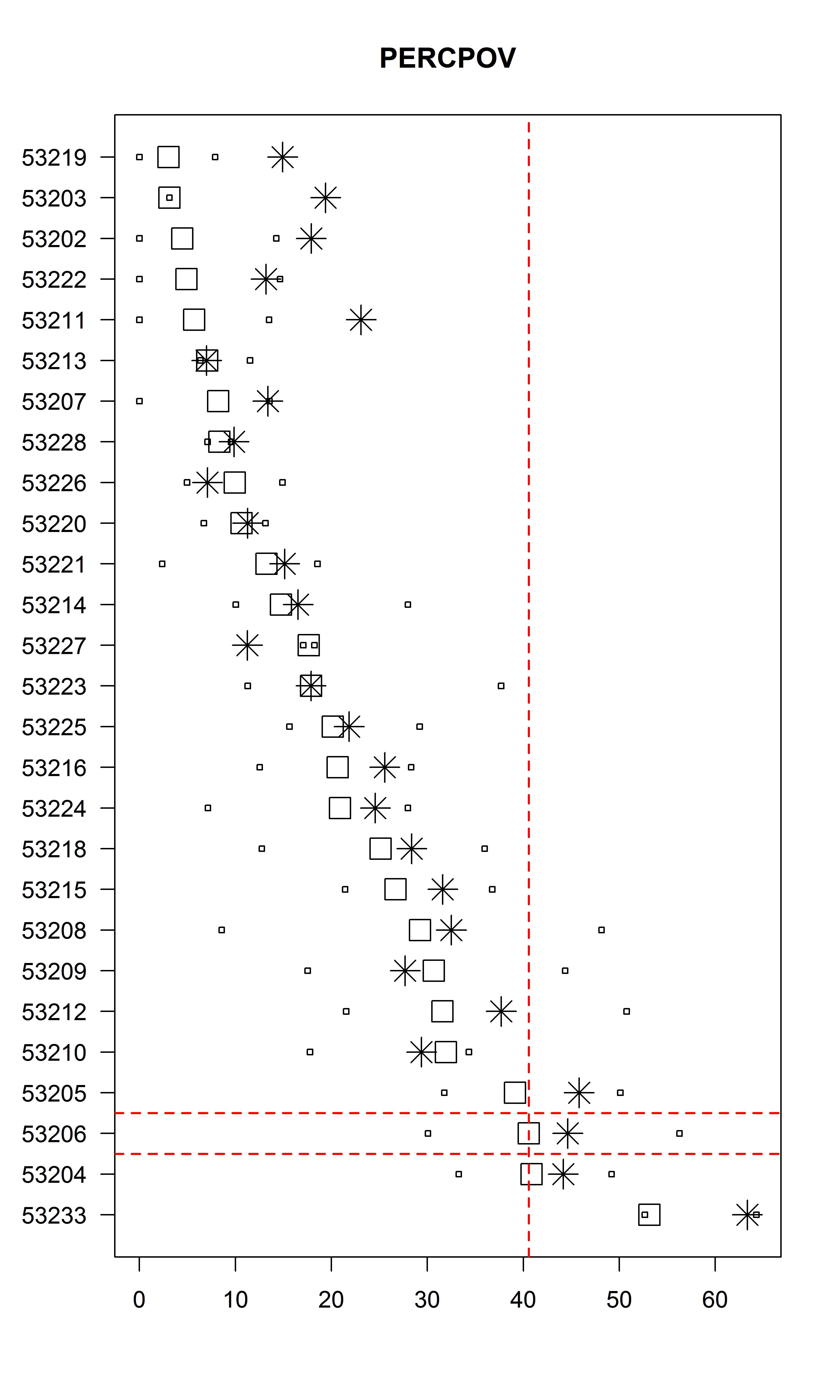}
\includegraphics[width=.5\textwidth]{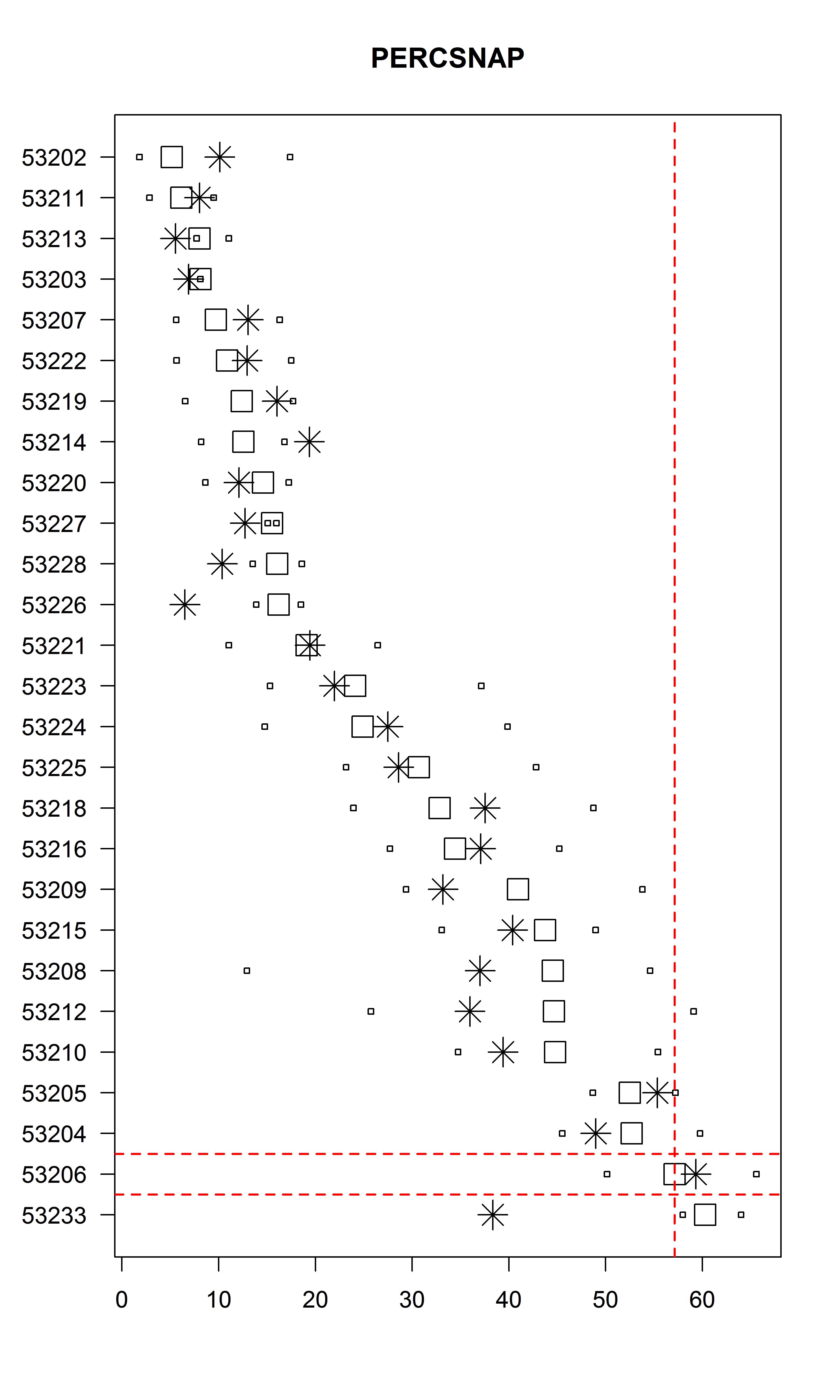}
\includegraphics[width=.5\textwidth]{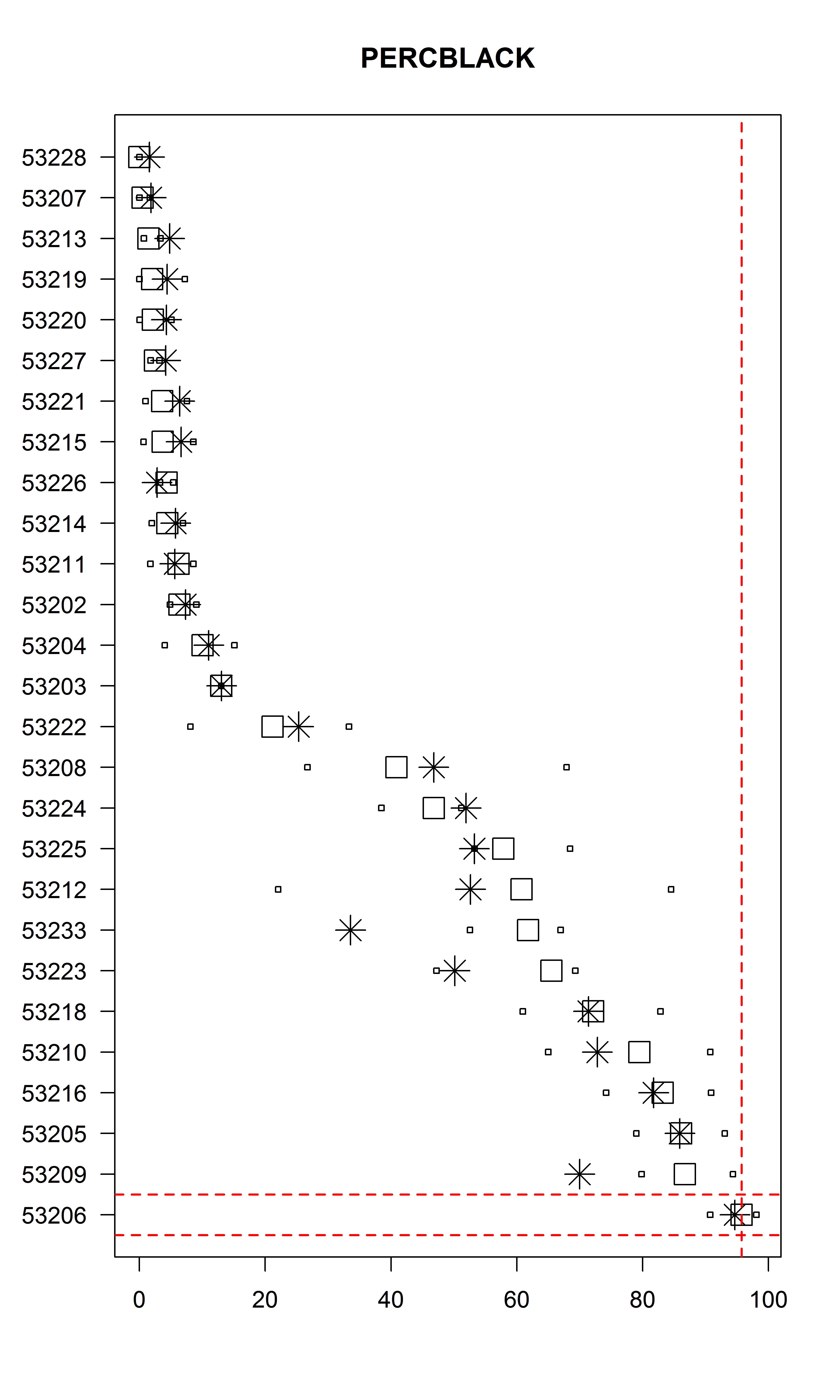}
\includegraphics[width=.5\textwidth]{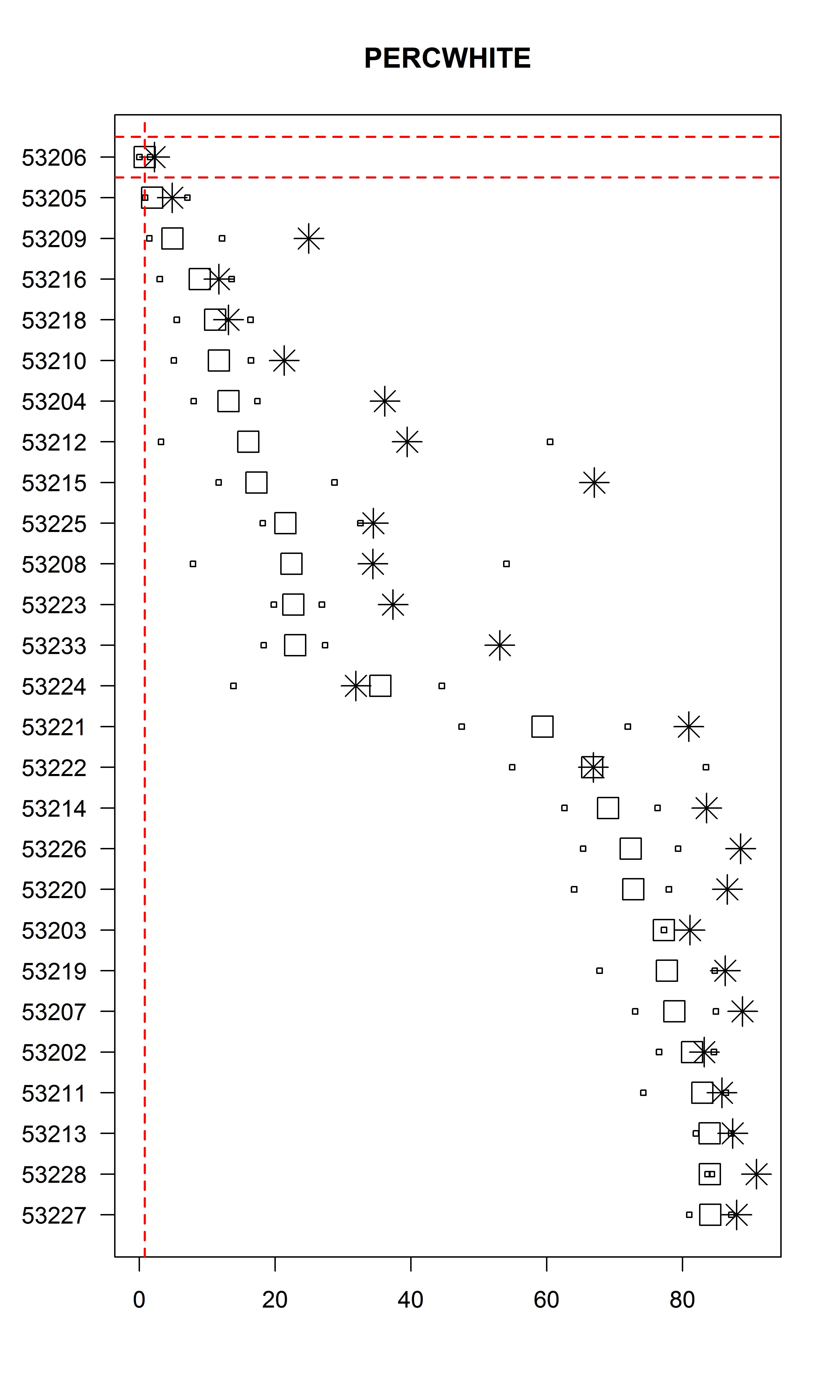}
\caption*{Note: $ \square$ = 25\%, 50\% (Median), 75\% values. Vertical line = 53206 median;\\ $\ast $ = ZIP code value.
}
\end{figure}

\begin{figure}[p]
\setcounter{figure}{1}
\hfill
\caption{Block-Group Distributions by ZIP Code, Continued.}
\includegraphics[width=.5\textwidth]{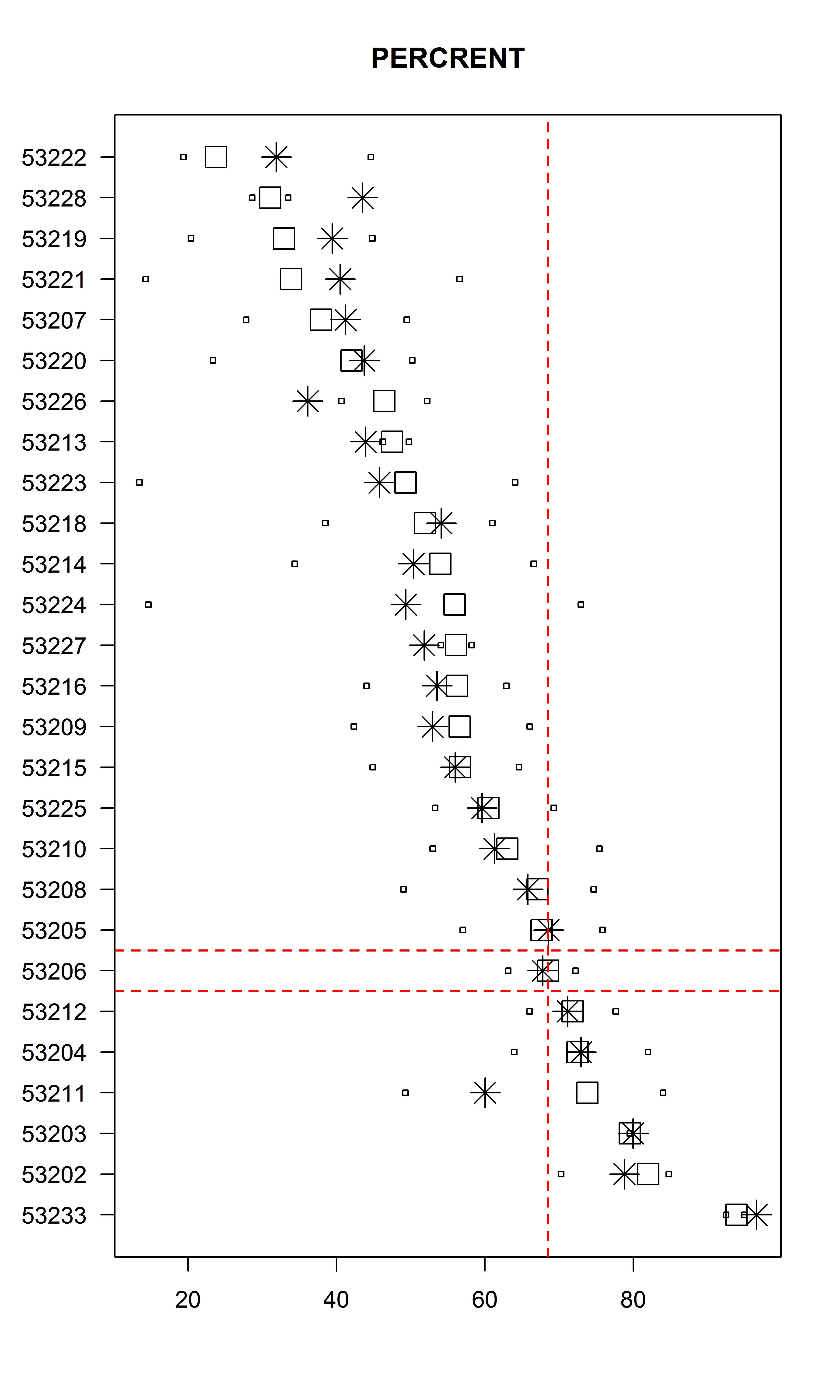}
\includegraphics[width=.5\textwidth]{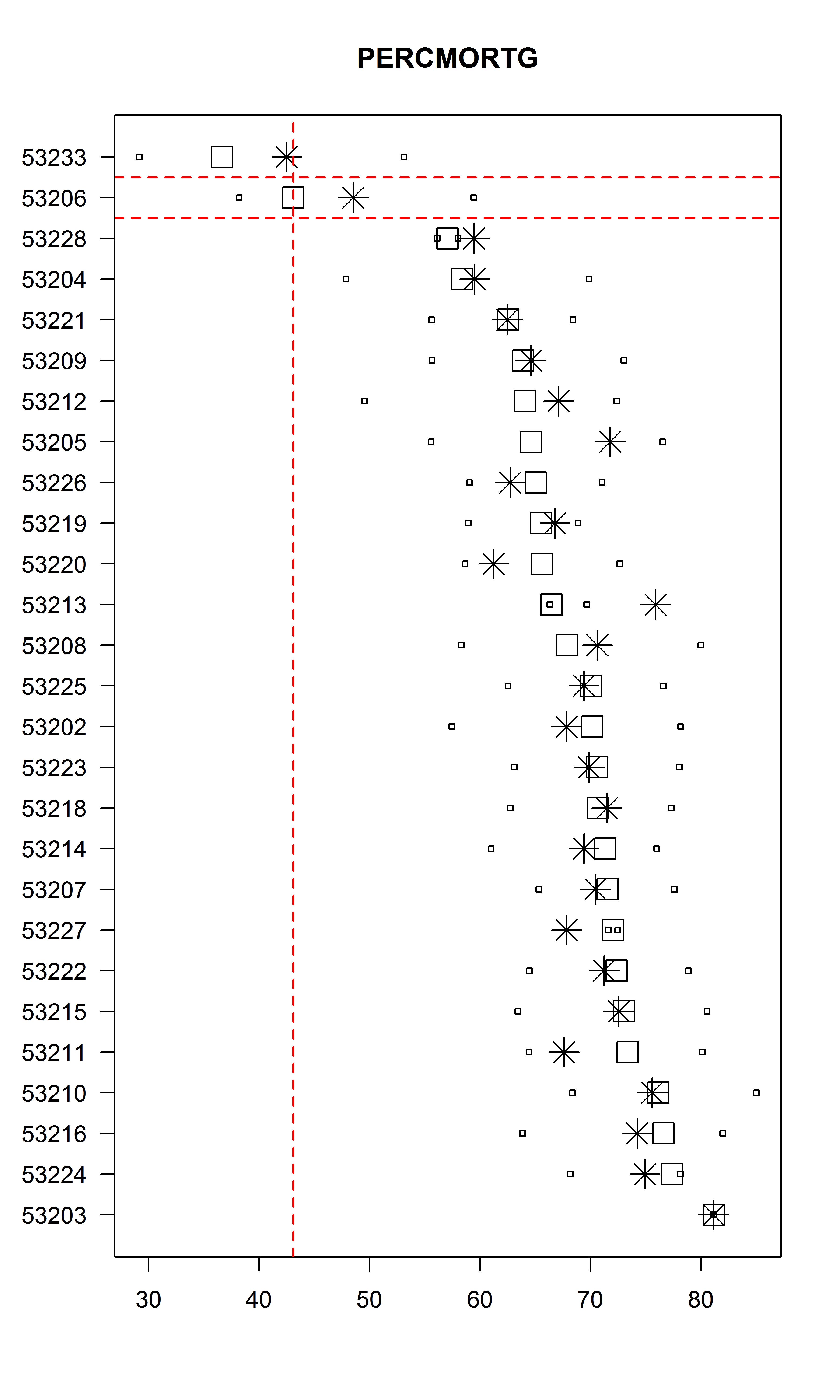}
\includegraphics[width=.5\textwidth]{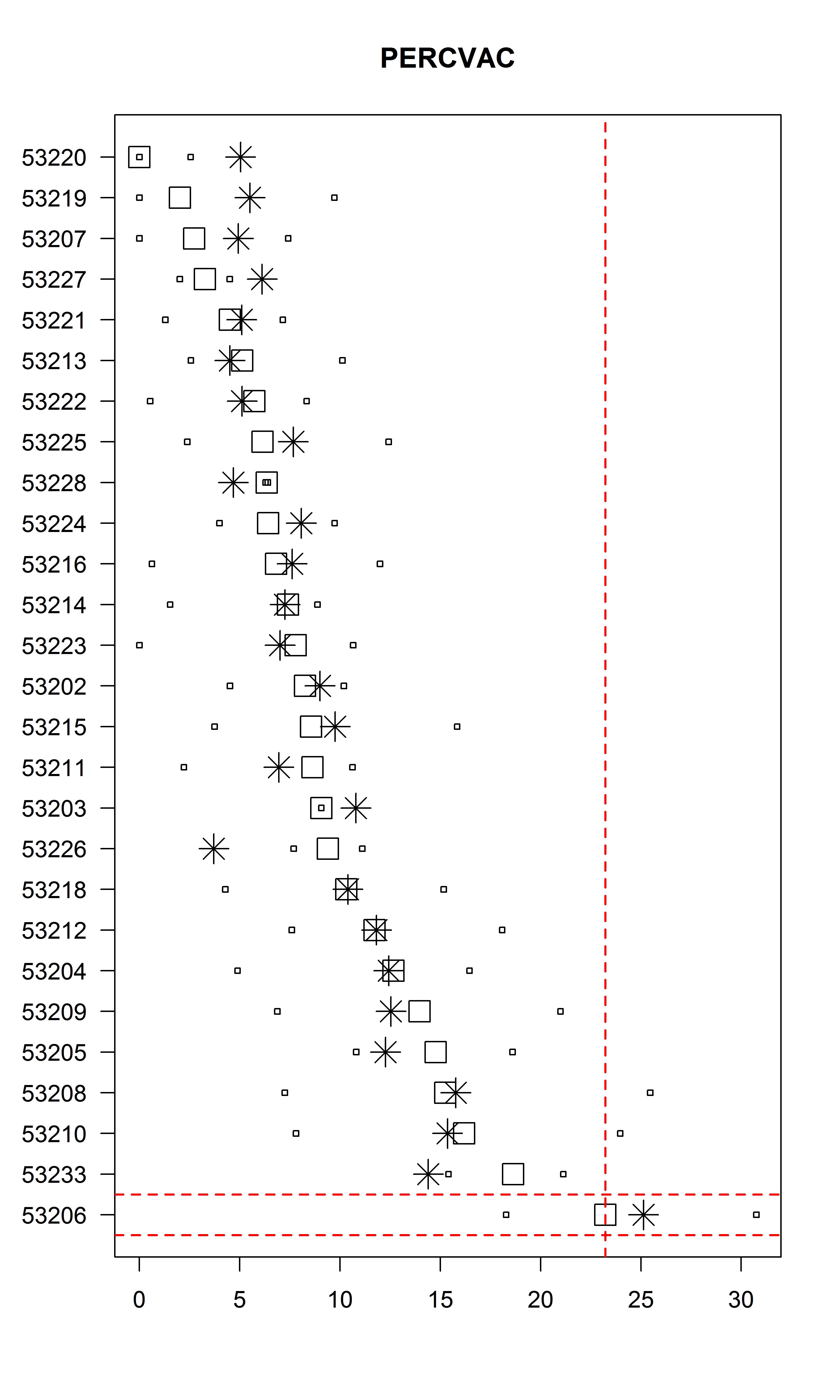}
\includegraphics[width=.5\textwidth]{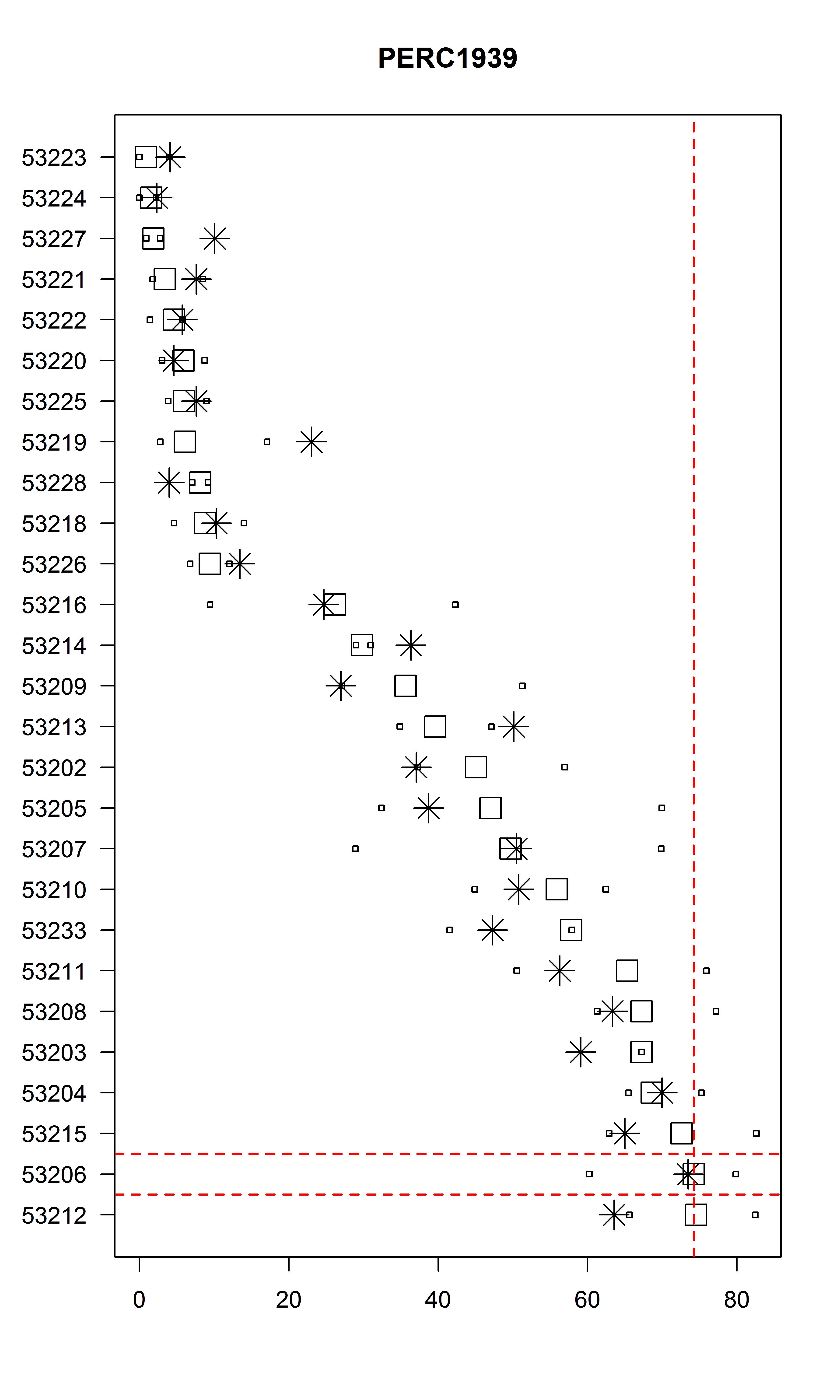}
\caption*{Note: $ \square$ = 25\%, 50\% (Median), 75\% values. Vertical line = 53206 median;\\ $\ast $ = ZIP code value.
}

\end{figure}
After we examine these statistics for the eight measures, we then use a subset of them to create a single index of socioeconomic conditions within each block group. We use Principal Component Analysis to capture the common variance among the chosen variables; this can be mapped and compared across the city. The resulting variable can also be used to compare Milwaukee with other cities nationwide. We generate index values for block groups in Chicago and Detroit and use them to compare 53206 with ZIP codes in these two cities. 

Finally, we compare this “53206” index with a separate index of deprivation. Used widely in the public health literature (see Carstairs, 1995), this measure incorporates a number of economic, education, housing, family, and other measures, and correlations are then calculated between the index and health status or other variables of interest. Hegerty (2019) calculates such a measure, and notes that this measure behaves differently in urban areas compared to suburban or rural ones, and captures different conditions than does the poverty rate alone. 

Following that approach, we apply PCA to capture the common variance of three variables:  the vacancy rate, which is used in the analysis above; as well as the unemployment rate and the percentage of adults above age 25 without a high school diploma. The first principal component of this set of variables is then mapped and compared to the five-variable index that was previously defined. While they are expected to track closely with one another, specific similarities, as well as differences can be examined in detail. In addition to estimating Spearman correlations between the two measures, we regress the PC5 “53206” index on the PC4 “Deprivation” index using two different methods. The first is traditional Ordinary Least Squares (OLS); the second is a spatial lag model in the vein of Hegerty (2019). This method incorporates “spatially lagged” variables—the values found in neighboring tracts—to capture spatial autocorrelation, or the effects of geographically proximate areas on conditions within each block group. Here, neighboring block groups are defined as Queen contiguity of order 1, or that only neighboring block groups that touch a given block group (even at a single point) are given weight. The rest receive a value of 0. The resulting weights matrix \textit{W} is included in the second equation.

\begin{equation}
PC5 = \alpha + \beta PC4 + \varepsilon
\end{equation}

\begin{equation}
PC5 = (I - \rho W)^{-1}(X\beta+\epsilon); X = PC4
\end{equation}

These results will allow us to examine alternative measures of socioeconomic conditions, both in the 53206 and across the city, while also keeping clustering and concentration in mind. Overall, we find that when spatial autocorrelation is taken into account, the regression model performs much better, and that the newly created “53206” index is closely related to the deprivation index. These and other findings are presented below. 

\section{Results}
Figure 2 shows the distributions for the three quantiles explained above, for each of the eight socioeconomic measures. One important finding is that the 53233 ZIP code, which includes a number of apartment buildings that were built later in the 2\textsuperscript{th} century, exceeds 53206’s values in a number of cases. Located just west of downtown, 53233 has the highest poverty rate and percentages of SNAP recipients in the city, although 53206’s 75\% threshold exceeds 53233’s median values. We also note that 53204 and 53205 also have high values; so does the poorest quarter of block groups in the 53212 ZIP code, which includes the relatively whiter and higher-income neighborhood of Riverwest. 

\begin{table}[ht]
\caption{Principal Components Analysis and Regression Coefficients.}
  \begin{adjustwidth}{-1.5in}{-1.5in}  
        \begin{center}
\begin{tabular}{lrlrrrrrlr}

&6-variable&&&5-variable&&3-variable&&\multicolumn{2}{l}{Regression  (DV = PC5)}\\
\hline

&EV&Variable&Loadings&EV&Loadings&EV&Loadings&Variable&beta\\
\hline
PC1&1.320&\%VAC&0.373&1.303&0.389&1.176&0.536&\%VAC&0.042\\
PC2&0.992&\%SNAP&0.507&0.948&0.520&0.909&0.627&\%SNAP&0.025\\
PC3&0.928&\%RENT&0.413&0.888&0.414&0.799&0.566&\%RENT&0.019\\
PC4&0.882&\%BLACK&0.381&0.816&0.402&&&\%BLACK&0.011\\
PC5&0.807&\%POV&0.490&0.703&0.496&&&\%POV&0.026\\
PC6&0.692&\%MORTG&-0.220&&&&&INPT&-3.498\\
\hline
\end{tabular}
        \end{center}
  \end{adjustwidth}
\end{table}

As noted in previous studies, 53206 has the lowest percentage of White residents and the highest percentage of Black residents; in this respect, the 53233 ZIP code is not even close. This reflects differences in location, housing stock, and history between the two parts of the city. The 53207 ZIP code is similar to 53206 in terms of percentage Black, and 53205 is similar in terms if percentage White, however.

\begin{figure}[ht]
\hfill
\caption{Block-Group Distributions of the Multivariate “53206” Index.}
 \makebox[\textwidth]{\includegraphics[width=.7\textwidth]{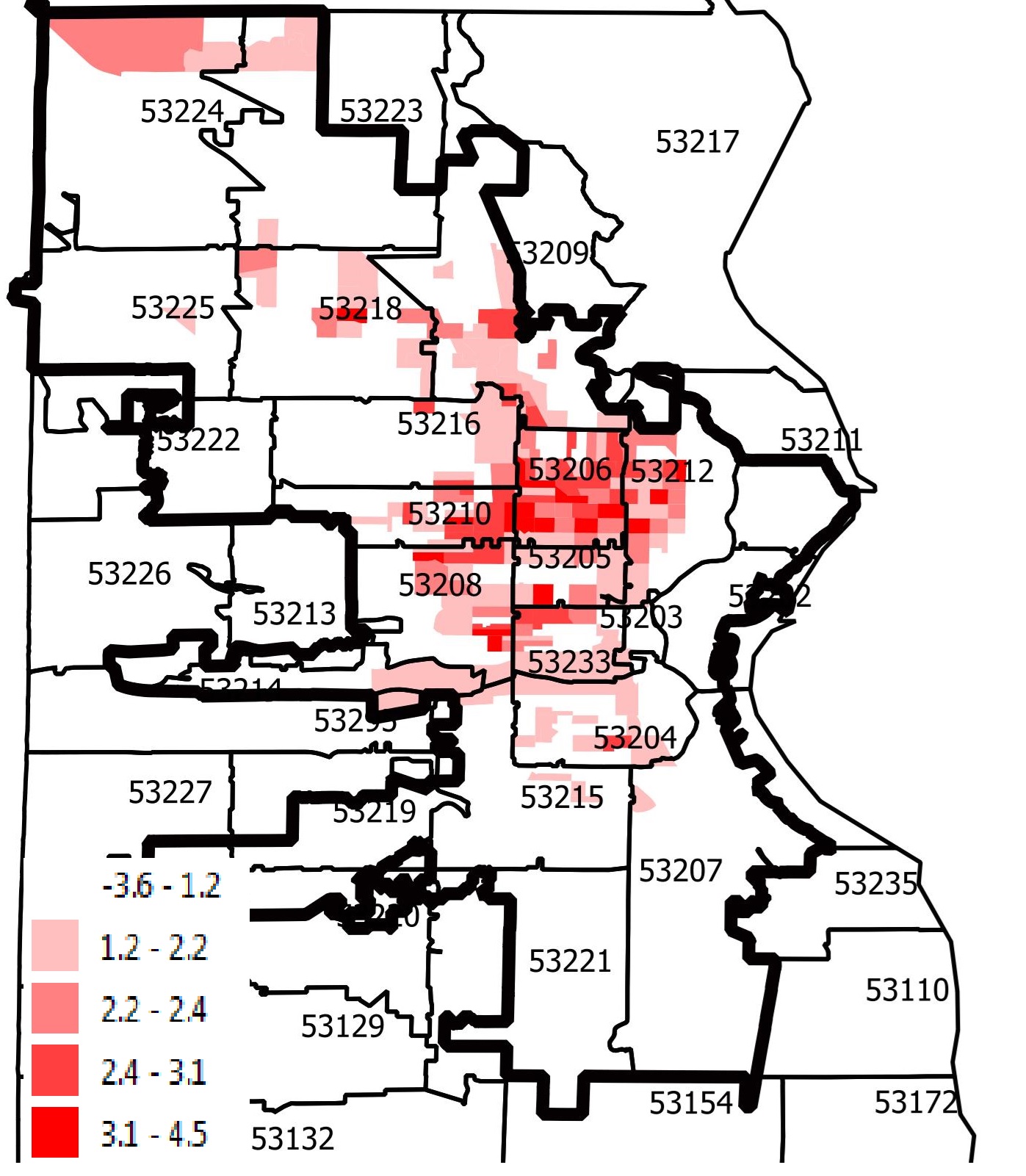}
\includegraphics[width=.5\textwidth]{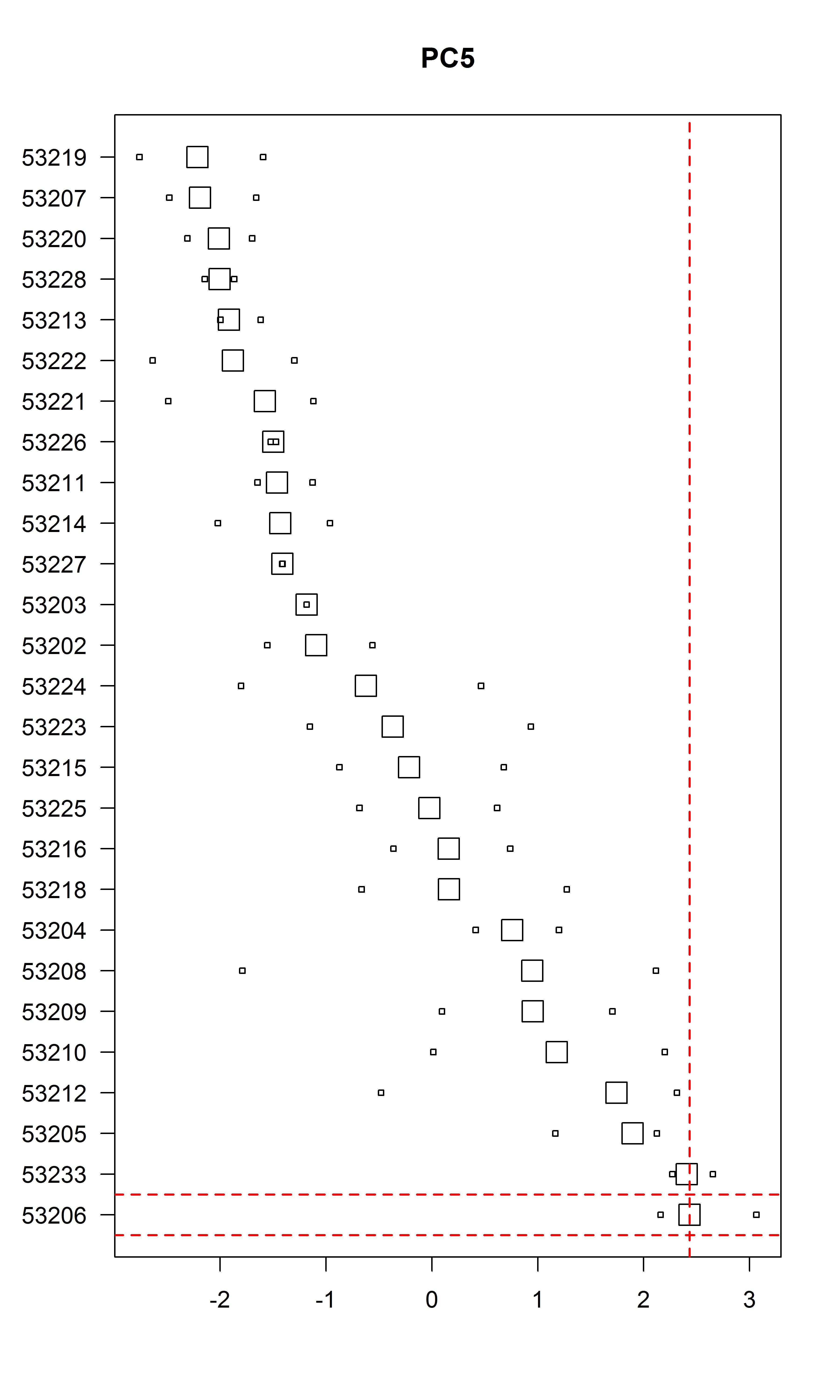}
}
\caption*{Map breaks = Quantile values.\\
$ \square$ = 25\%, 50\% (Median), 75\% values. Vertical line = 53206 median.
}

\end{figure}

While the 53233 ZIP code has the largest proportion of renters, 53206 has the lowest proportion of mortgages. This suggests that financial exclusion—particularly limited access to credit—may be a particularly sever issue in this part of the city. In fact, 53206 ranks seventh in terms of the percentage of renters. Perhaps some homeowners in this older part of the city, who likely bought their properties decades ago, have been able to pay off their mortgages. Or, as mentioned above, comparatively low property values might encourage cash transactions. But these results—whether positive or negative—should be studied further. The 53206 ZIP code also has the city’s highest median vacancy rate; only the upper quartiles of 53208 and 53210 exceed the 53206 median value. Multiple ZIP codes have upper-quartile poverty rates and proportions of SNAP recipients that exceed the 53206 median as well. 

To capture the common variance among our set of variables, we use the first principal component of each subset. The results of this analysis, for sets of six, five, and three variables, are presented in Table 3. Only one eigenvalue in each case exceeds 1.0, signifying the presence of a single principal component for each subset. For each choice of variables, the proportion of SNAP recipients loads highest on the resulting index. Of the three alternatives, we choose the five-variable index, choosing to exclude the percentage of households with mortgages. As noted above, there numerous other explanations for this phenomenon that might capture different economic processes than those we set out to explore here. We name the resulting variable PC5, and also refer to it as the “53206” Index. Table 3 also provides the coefficients from a regression of PC5 on each of its components; these can be used with other cities’ data to create the index using fitted values.

\begin{figure}[ht]

\hfill
\caption{Block-Group Distributions of the Multivariate “53206” Index:\\ Detroit (Left) and Chicago (Right).}
 \noindent
 \makebox[\textwidth]{\includegraphics[width=1.3\textwidth]{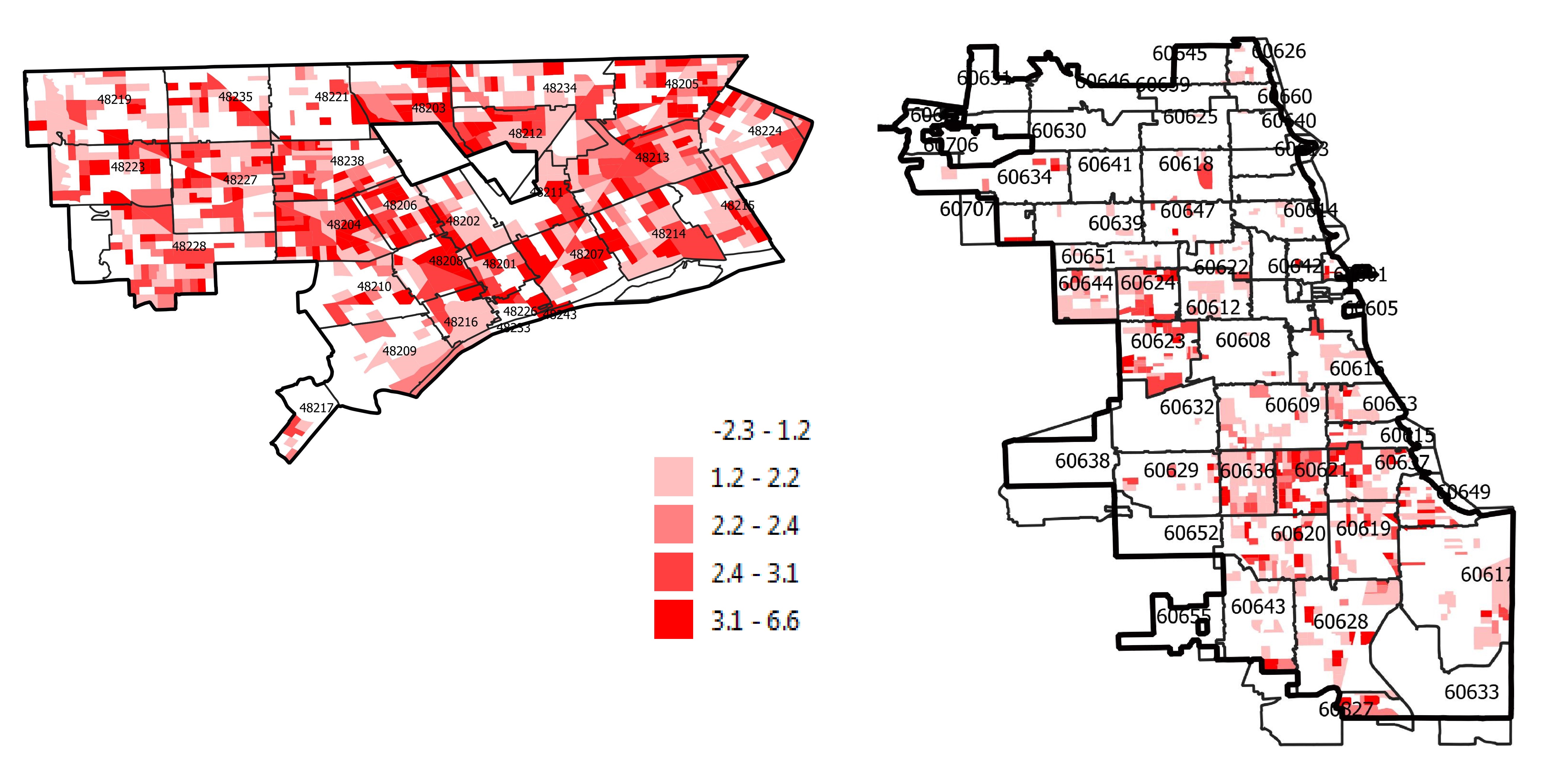}}%

\caption*{Color breaks based on Milwaukee values.}
\end{figure}

Figure 3 shows the distributions of the multivariate “53206” index. Spatially, it is clustered among block groups in this ZIP code, as well as in contiguous areas. As was the case for the other variables, parts of the 53210 and 53212 ZIP codes can be considered to be part of the 53206 cluster. Any programs specifically targeted to this area should consider crossing these arbitrary lines. The quantile distributions show that 53206 block groups have the highest median and 75\% index values, with the 53233 median similar to 53206’s. The top 25\% of block groups in the  53205, 53208, 53210, and  53212 ZIP codes are also high, but do not exceed the 53206 median value. In this sense, then, Milwaukee’s 53206 stands out among the other city ZIP codes.

\begin{figure}[p]
\hfill
        \begin{center}
\caption{Block-Group Distributions of the Multivariate Index Within Each ZIP Code: Detroit (Left) and Chicago (Right).}
\noindent
 \makebox[\textwidth]{
\includegraphics[width=.7\textwidth]{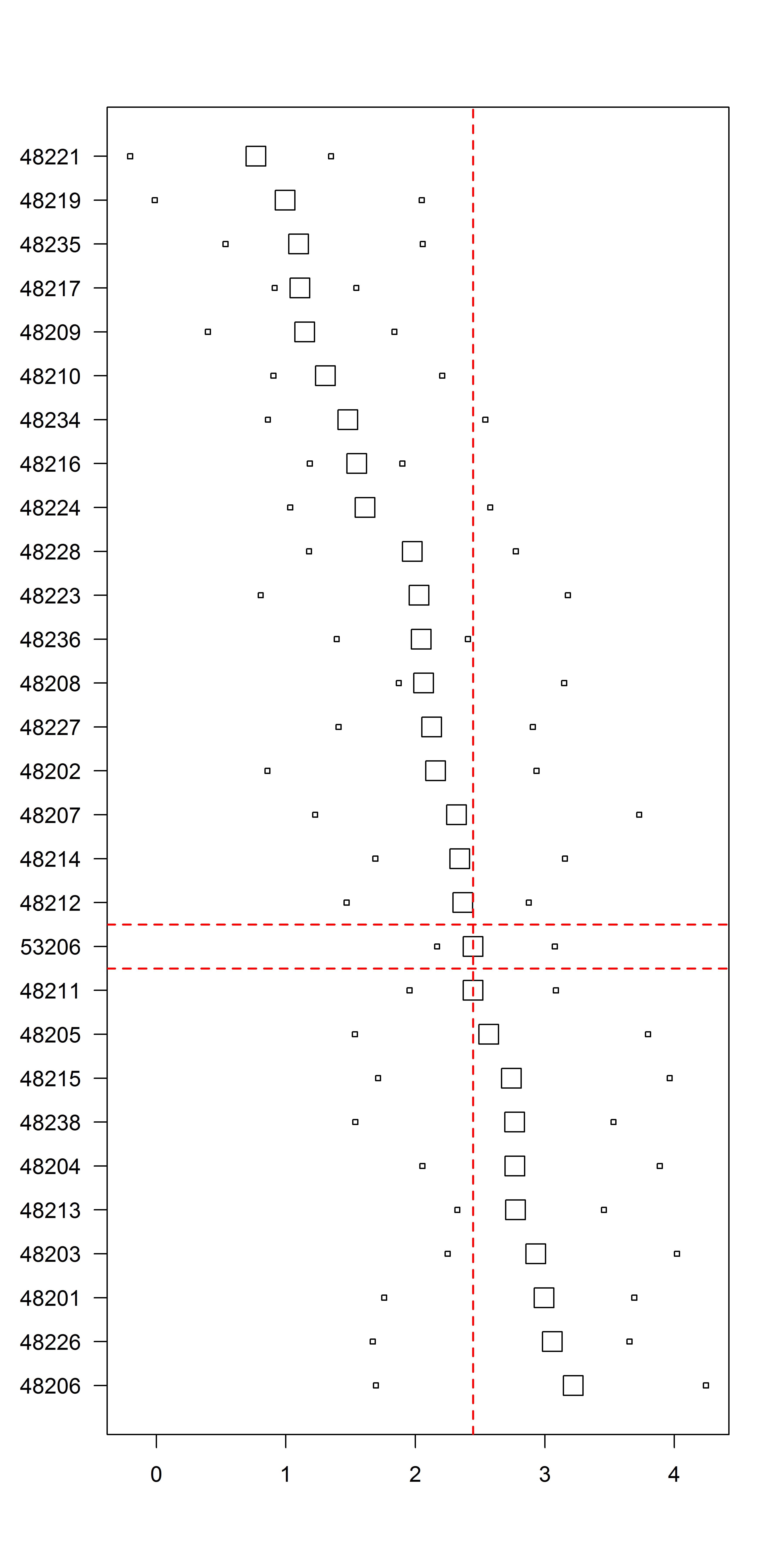}
\includegraphics[width=.7\textwidth]{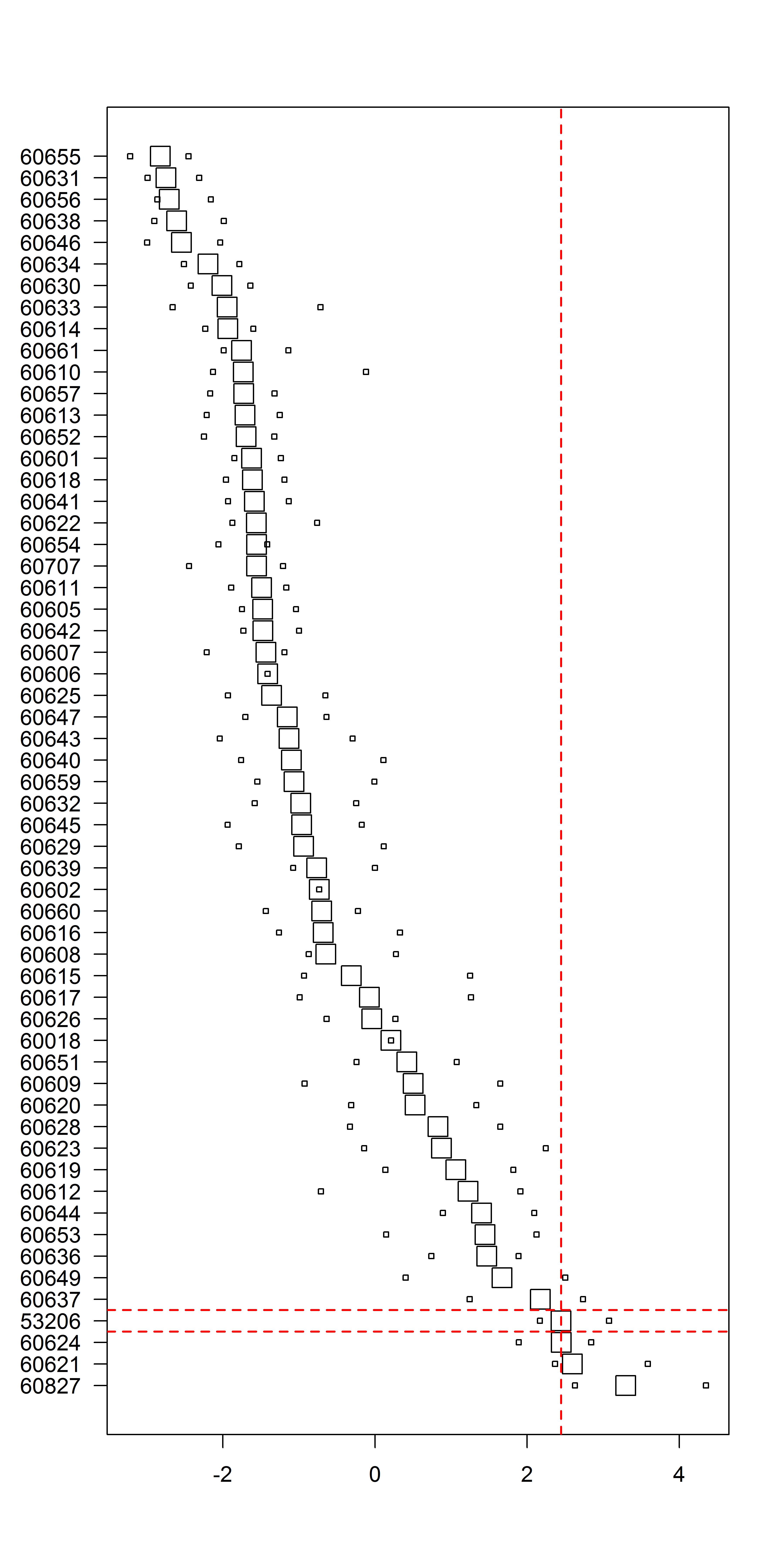}}
\caption*{Note: $ \square$ = 25\%, 50\% (Median), 75\% values. Vertical line = 53206 median.
}

\end{center}
\end{figure}

How would this area compare against ZIP codes elsewhere in the country? We recreate the 53206 index by constructing the fitted values using the regression coefficients from Table 3 and similar Census block group data for Chicago and Detroit. Both have reputations for their severity of urban conditions within their borders. Figure 4 depicts the spatial distribution of this index, with category breaks based on the 10, 25, 50, and 75 percent quantile values of the 53206 block groups. High index values are clustered across Detroit, but are more concentrated on West and South Sides of Chicago. While high values within these two areas might be expected, we also find them outside these clusters, particularly on the Northwest Side and in Rogers Park.  

Figure 5 presents the within-ZIP-code distributions for this index in both cities. Milwaukee’s 53206 would be near the middle among Detroit ZIP codes, but would be in third place among Chicago’s 58 ZIP codes examined here. This provides useful context when comparing neighborhoods or cities elsewhere that are often “written off” because of poverty, crime rates, or even racial makeup.

\begin{figure}[ht]
\hfill
\caption{Deprivation Index (PC4) in Milwaukee Block Groups and ZIP Codes.}
 \makebox[\textwidth]{\includegraphics[width=.7\textwidth]{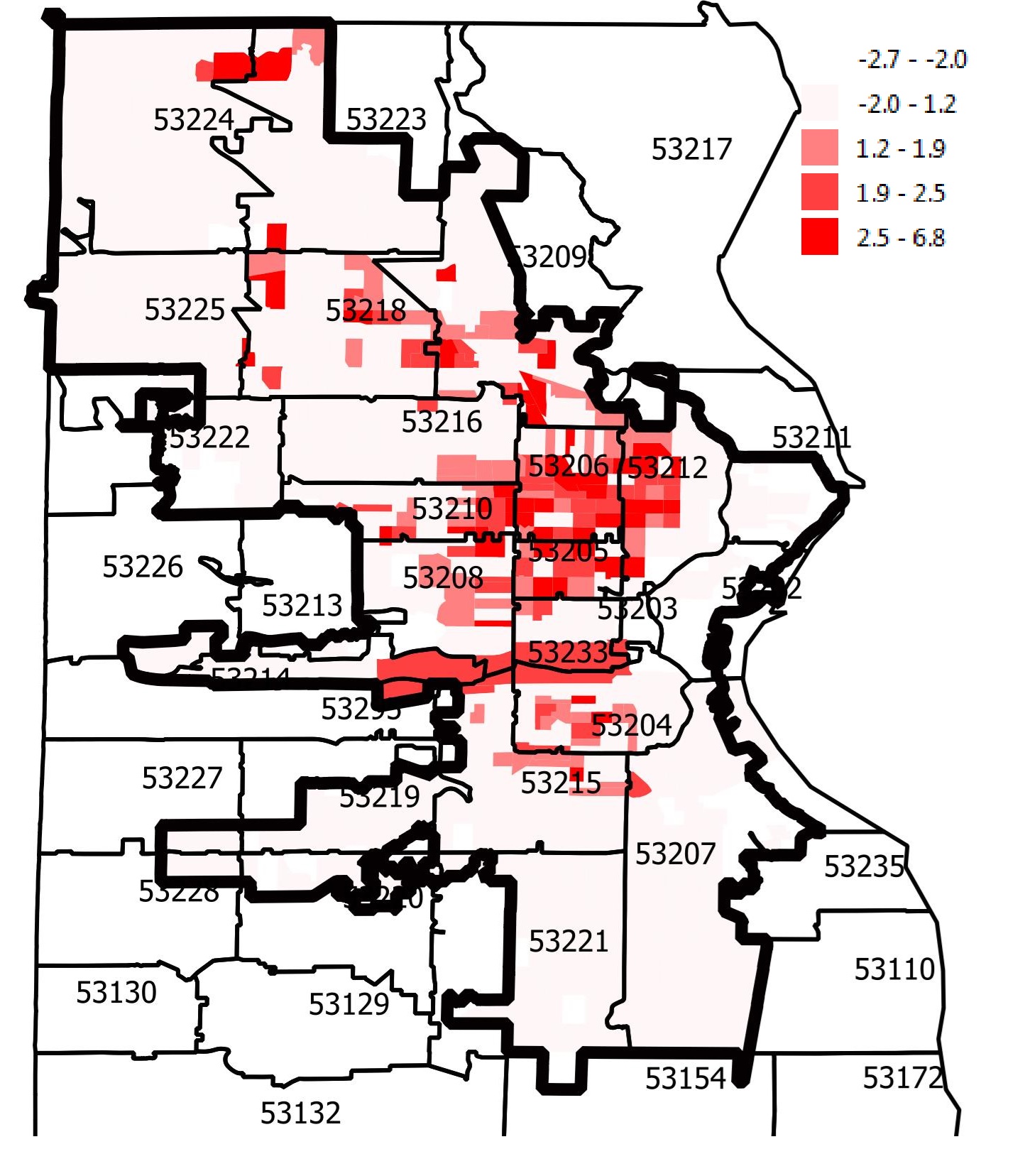}
\includegraphics[width=.5\textwidth]{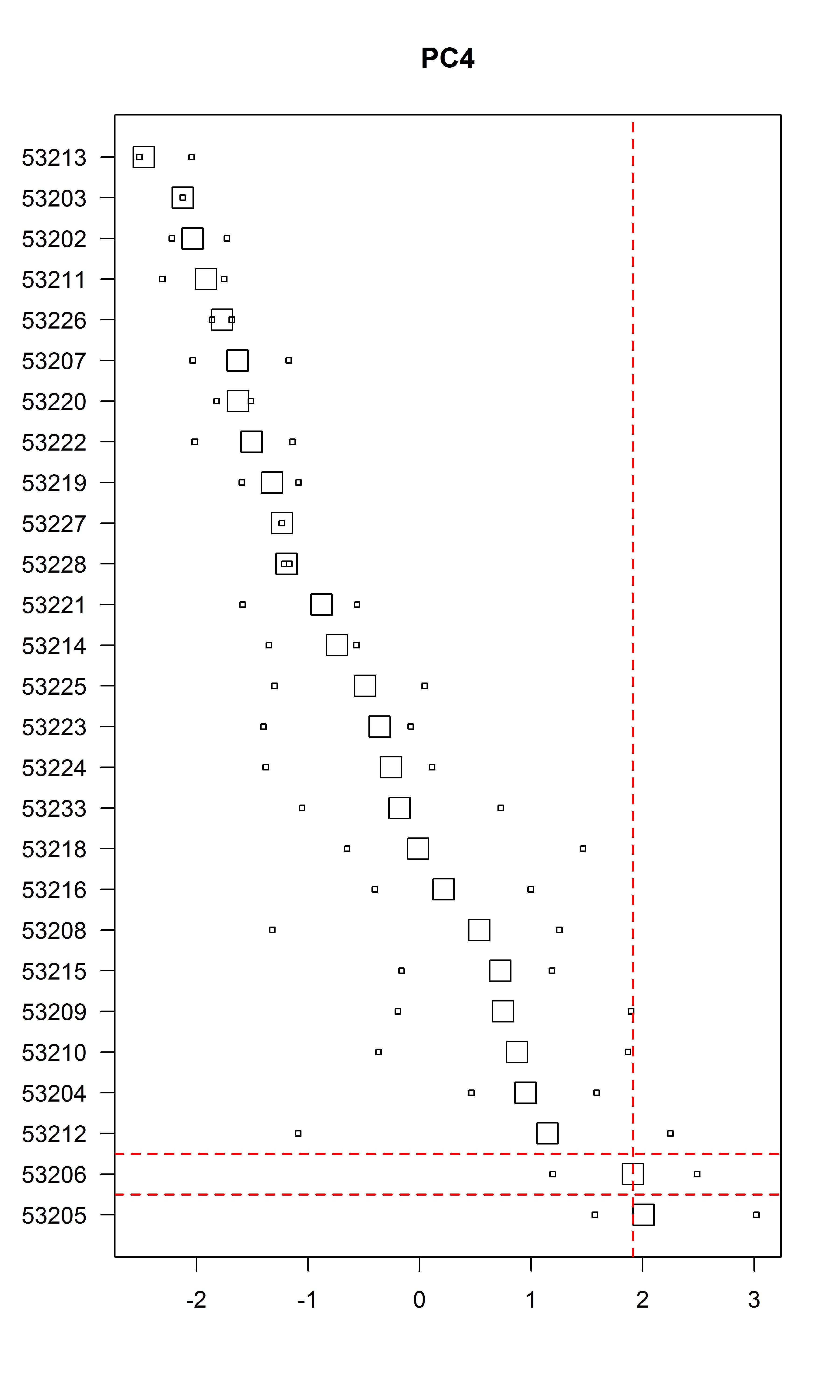}
}
\caption*{Map breaks = Quantile values.\\
 $ \square$ = 25\%, 50\% (Median), 75\% values. Vertical line = 53206 median.
}

\end{figure}

Finally, we compare the five-variable “53206” index with the more general, multivariate index of socioeconomic deprivation tested by Hegerty (2019). These are closely related to one another, with a Spearman correlation of 0.807. The distributions of this index are given in Figure 6. While the index clusters appear to be similar to the other index, one major difference is that Milwaukee’s 53205 ZIP code is shown to have the highest median and 75\% deprivation values. Clearly there is some difference between the two measures, which our new index captures.  

\begin{table}[ht]
\caption{Regression Results (PC5 Regressed on PC4).}
        \begin{center}
\begin{tabular}{lrr}

&Ordinary Least Squares&Spatial Lag Regression\\
&Coefficient (s.e.)&Coefficient (s.e.)\\
\hline
INPT&0.013 (0.044)&-0.016 (0.343)\\
Deprivation&0.812 (0.027)&0.460 (0.028)\\
$\rho$&&0.570\\
R\textsuperscript{2}&0.622&\\
\hline
\end{tabular}
\caption*{Standard errors in parentheses.}
        \end{center}
\end{table}

\begin{figure}[ht]
\hfill
 \begin{center}
\caption{“53206 Index” Values (PC5) Values vs. Deprivation Index (PC4) in Milwaukee Block Groups.}
\includegraphics[width=1\textwidth]{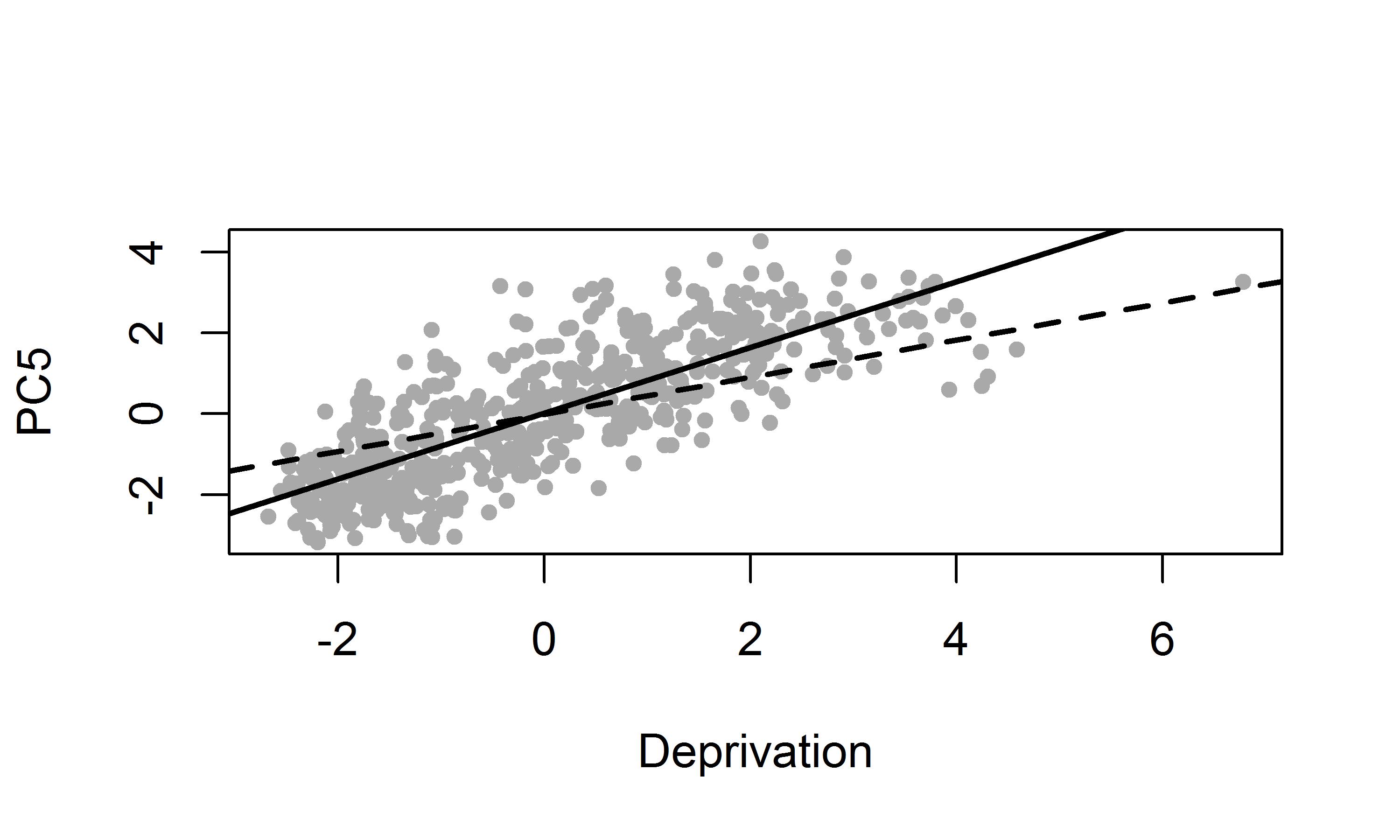}
\caption*{Solid line = OLS regression line; Dashed line = spatial lag regression line. 
}
\end{center}
\end{figure}

\begin{figure}[ht]
\hfill
\caption{Regression Residuals by ZIP Code (with 2-standard-deviation bands).\\Left = OLS Model; Right = Spatial Lag Model}
 \makebox[\textwidth]{
\includegraphics[width=.7\textwidth]{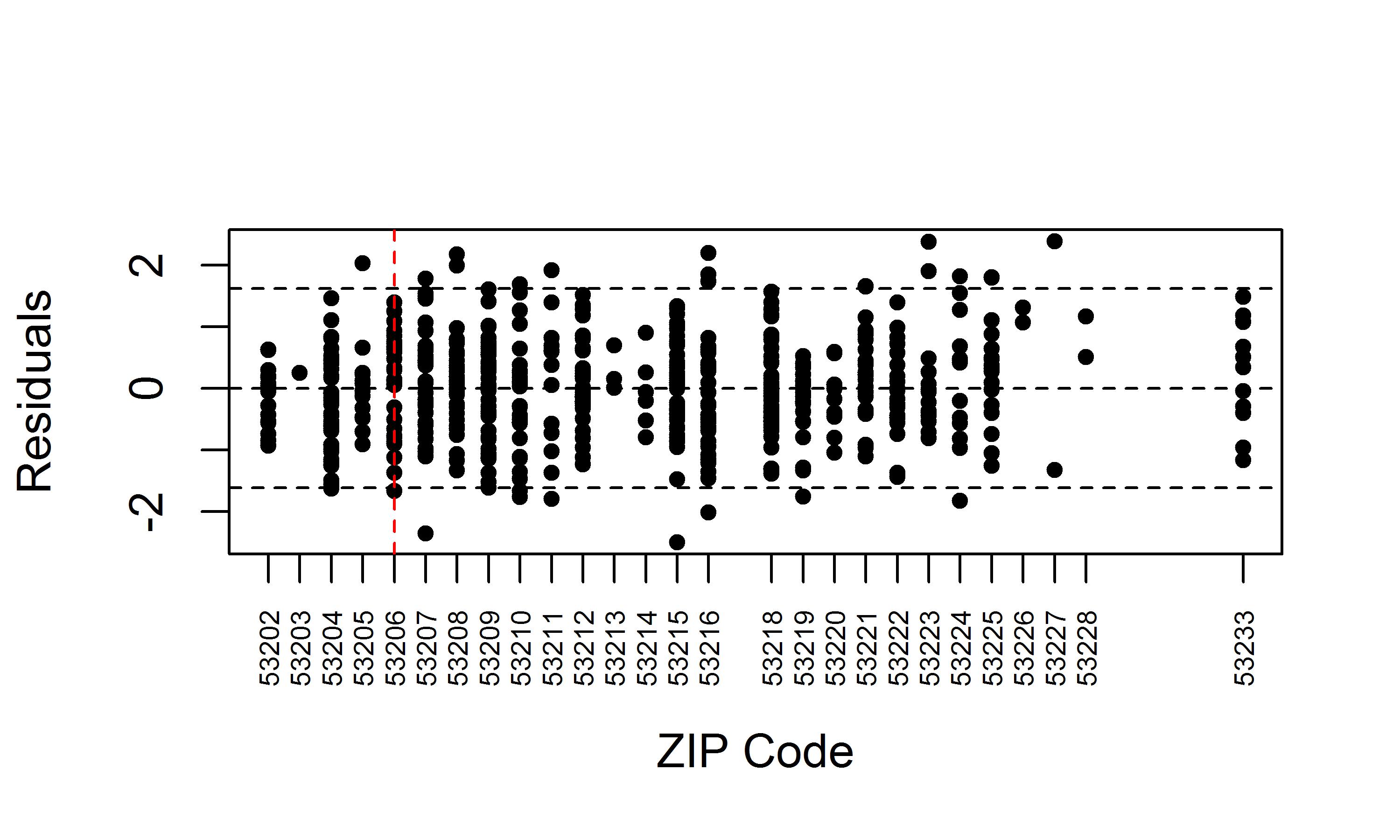}
\includegraphics[width=.7\textwidth]{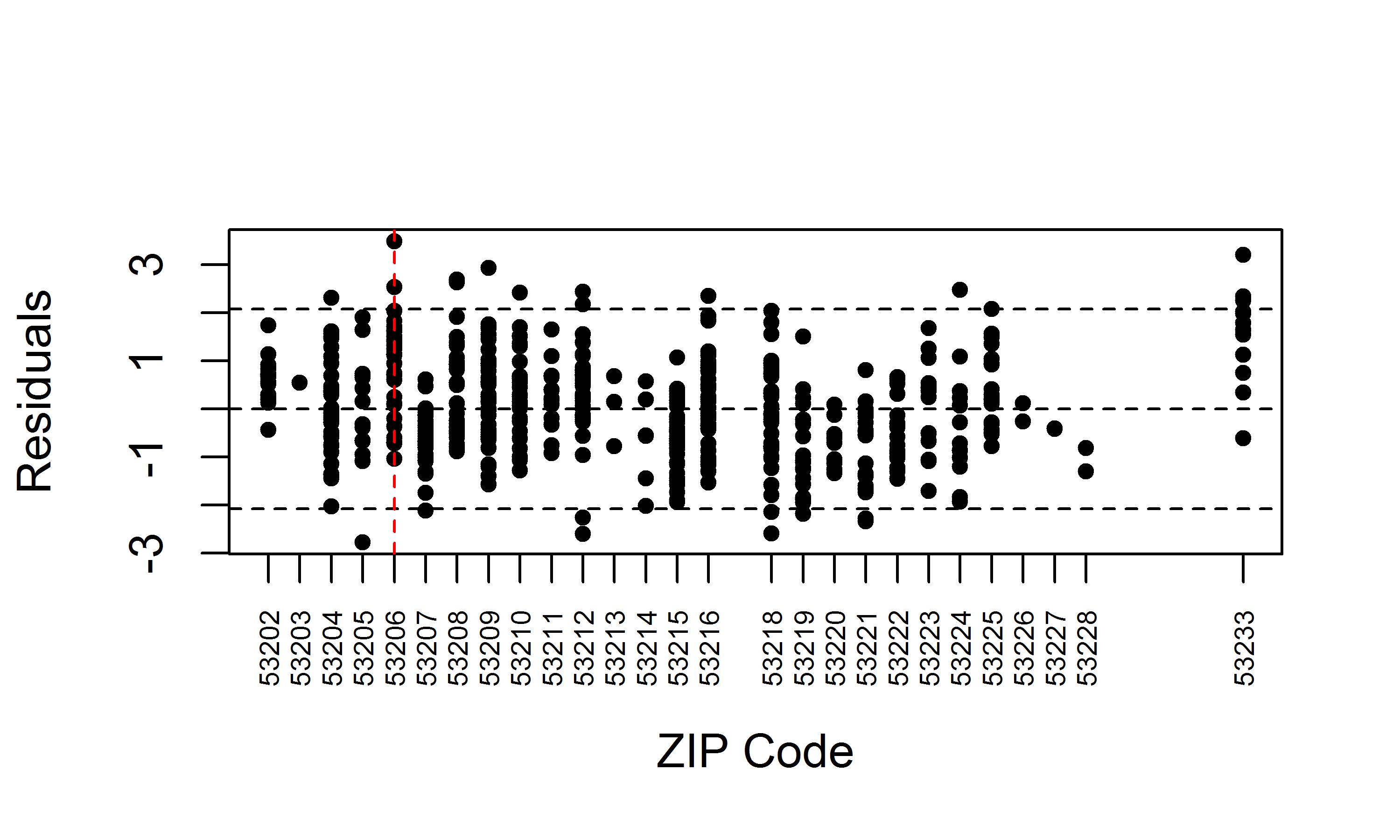}

}
\end{figure}

Table 4 presents the results of a bivariate regression of the “53206” index (PC5) on socioeconomic deprivation (PC4). There is a significantly positive relationship between the two, although the effect is smaller when spatial autocorrelation is taken into account. Most likely the fact that index values are clustered plays an important role in making the 53206—and contiguous areas of neighboring ZIP codes—so unique in their socioeconomic conditions. 

Figure 7 compares individual values for each index. There are a handful of high-deprivation block groups with relatively low “53206” index values. These outliers are captured by the spatial lag regression. Regression residuals for both estimation methods are also depicted in Figure 8. These are grouped by ZIP code.

Figure 8 presents the residuals from both models. There are large, positive residuals in the OLS model for the 53206 and 53233 ZIP codes, suggesting that the “53206” index often exceeds the values predicted by the deprivation index in these key areas. This effect disappears when spatial autocorrelation is incorporated into the model, however. We can conclude, therefore, that clustering needs to be taken into account when addressing deprivation in these neighborhoods, but that overall, the general deprivation index captures the same effects as does the specialized “53206” index.

\section{ Conclusion}
While much of the city suffers from a number of socioeconomic challenges, Milwaukee’s 53206 ZIP code has received particular attention from academics, politicians, and even filmmakers. While there is need within this specific areas, focusing on a single ZIP code—or ZIP codes in general—might neglect similar areas elsewhere in the city. This study examines the 551 Census block groups in Milwaukee, comparing a set of socioeconomic variables for these small units, as well as for their corresponding ZIP codes. We also disaggregate these corresponding block groups by calculating the median, as well as the 25 and 75 percent quartile values for each variable. We find while their overall average is often lower than 53206’s, many contiguous block groups have characteristics similar to this ZIP code; their top- or bottom- quartile values often are similar to the 53206 median. In addition, the 53233 ZIP code has conditions that are often even more extreme than what is the case in 53206.   

Of the eight variables we consider in our analysis, we combine five to create a unique “53206” index. We find that this ZIP code would still be among the most extreme if located in Chicago, but that it would be nearly average if located in Detroit. We compare this new index to a more traditional deprivation index, and find that while it may capture certain unique neighborhood effects, the new index is similar to the alternative when spatial effects are taken into account. The alternative index, however, finds the 53205 ZIP code to have the highest levels of deprivation in the city.

Overall, therefore, we conclude that Milwaukee’s 53206 ZIP code is not unique—either in Milwaukee itself, or compared to other cities. The 53233 and 53205 ZIP codes have similar levels of deprivation and other economic stresses, and parts of Chicago—and a much larger proportion of Detroit—are even more extreme. Further study could extend this analysis for other cities. Examining block groups also finds areas on the other side of the arbitrary ZIP code border to be no different from 53206.  Policymakers and community members, while they are commendably drawing attention and resources to this area of concentrated deprivation, should not lose sight of other neighborhoods nearby.  

\nocite{Hegerty2019}
\nocite{Carstairs1995}
\nocite{Wilson200809}
\nocite{Massey1988}
\nocite{Hegerty2017}
\nocite{Jargowsky1996}
\nocite{Lockwood2007}
\nocite{Ginsberg2011}
\nocite{Ward2008}
\nocite{Openshaw1983}
\nocite{McQuirter2016}
\nocite{Spicuzza2019}
\nocite{Levine2019}

\bibliographystyle{abbrv}
\bibliography{53206Paper}

\end{document}